\documentclass[epj]{svjour}
\usepackage{graphicx}
\usepackage{placeins}
\usepackage{hhline}
\usepackage{slashed}
\usepackage{xcolor}
\usepackage{amssymb, amsmath}

\begin{document}

\title{Tetraneutron condensation in neutron rich matter}
\titlerunning{Tetraneutron condensation in neutron rich matter}

\author{O. Ivanytskyi, M. \'Angeles P\'erez-Garc\'ia and  C. Albertus}%
\institute{Department of Fundamental Physics, University of Salamanca, Plaza de la Merced S/N E-37008, Spain
}%




\date{Received: date / Revised version: date}

\abstract{
In this work we investigate the possible condensation of tetraneutron resonant states in the lower density neutron rich gas regions inside Neutron Stars (NSs). Using a relativistic density functional approach we characterize the system containing different hadronic species including, besides tetraneutrons, nucleons and a set of light clusters ($^3$He, $\alpha$ particles, deuterium and tritium). $\sigma,\omega$ and $\rho$ mesonic fields provide the interaction in the nuclear system. We study how the tetraneutron presence could significantly impact the nucleon pairing fractions and the distribution of baryonic charge among species. For this we assume that they can be thermodynamically produced in an equilibrated medium and scan a range of coupling strengths to the mesonic fields from prescriptions based on isospin symmetry arguments. We find that tetraneutrons may appear over a range of densities belonging to the outer NS crust carrying a sizable amount of baryonic charge thus depleting the nucleon pairing fractions. 
} 

\maketitle




\section{Introduction} 
The essential identity of nuclear forces acting between nucleons lies behind the idea of the possible existence of resonant tetraneutron states (${}^4n$). Long standing searches of this four neutron state were initiated by studying its possible formation in the break-up reaction ${}^{14}Be\rightarrow{}^{10}Be+{}^4n$ \cite{4n_exp1}. Despite the fact that the existence of bound tetraneutrons was theoretically excluded with high confidence \cite{4n_th1,4n_th2,4n_th3,4n_th4,4n_th5,4n_th6,4n_th7}, no arguments able to rule out the resonant ${}^4n$ state were presented. A new splash of enthusiasm to studies of the four neutron resonance is motivated by a recent experimental claim that such a state with an excitation energy $E_{4n}=0.83$ MeV and width $\Gamma_{4n}=2.6$ MeV was detected in the exchange reaction $\alpha+{}^8He\rightarrow{}^8Be+{}^4n$ \cite{4n_exp2}. Intensive theoretical studies confirmed that tetraneutrons can exist in the form of resonances. In Ref. \cite{4n_th8} and  in what can be assumed to be vacuum conditions its excitation energy and width were predicted to be $E_{4n}=0.8$ MeV and $\Gamma_{4n}=1.4$ MeV, respectively. Larger values of these quantities of about $E_{4n}=$7.3 MeV and $\Gamma_{4n}=3.7$ MeV were also reported in \cite{4n_th9}. Other robust theoretical approaches predict even larger widths of tetraneutrons \cite{4n_th1,4n_th6,4n_th10}.

Such large widths seem to indicate that tetraneutrons are not likely to exist under the form of stable aggregates although this may be possible if confined by strong external fields \cite{4n_th1,4n_th8}. The simplest system exhibiting such a tetraneutron confinement could be associated to ${}^8He$. In such a system, nuclear forces create a significant potential barrier which prevents the break up of tetraneutrons and confines them around $\alpha$-cores. Along the same line, it may be of interest to investigate whether in an extended system, such as an strongly interacting nuclear matter medium, the existence of these resonances could be due to the presence of strong mesonic fields mediating nucleon-nucleon forces. Such an effect could be crucial for the description of neutron rich matter, where the probability of creation of a ${}^4n$ resonance could be, in principle,  statistically favoured. If the fraction of such resonant states was high enough they could significantly affect the properties of $\beta$-equilibrated nuclear matter even despite their associated short lifetime.
NSs seem to be suitable places to help shed light on the possible existence of tetraneutrons. As we will argue later in the manuscript, these particles can populate the stellar medium with an onset density that can be well below nuclear saturation density.  The main question we address in this contribution is, thus, twofold. First,  whether tetraneutrons could exist in  a neutron rich medium and second, how they could affect its microscopic properties.

The possible existence of tetraneutrons in a nuclear medium is linked to their resonant character \cite{4n_th2,4n_th3,4n_th4,4n_th5,4n_th6,4n_th7,4n_th10,4n_th11}. 
An effective treatment based on non-fundamental hadronic degrees of freedom has proven to be successful when describing the interior of NSs at intermediate energies.  Therefore, in this work we will consider the appearance of tetraneutrons in $\beta$-equilibrated nuclear matter by using  a relativistic field theoretical model based on, besides this new species, baryons, leptons, light clusters and mesonic degrees of freedom in a similar fashion to that done in previous works \cite{RMFclusters1,RMFclusters2,RMFclusters3,RMFclusters4}. Further,  such a treatment is also analogous to that followed for dibaryons in \cite{RMFdibaryons1,RMFdibaryons2}. Additional work on the presence of other resonances has also been performed, for example in \cite{Delta1} where the effects of the mass distribution and the associated mass-dependent lifetimes of $\Delta(1232)$ resonances were partially considered, although a full in-medium treatment is still missing. Further works on this include  \cite{p1,p2,d2}.
It is important to remark that for low densities further corrections from correlations beyond a relativistic mean field (RMF) treatment are needed to precisely characterize the system. The inhomogeneous phases yield an optimized and lower free energy when compared to that of homogeneous matter. This determines what heavier clusters (nuclei) appear in the system at fixed temperature, density and proton fraction. Some approaches obtain the actual matter configuration analyzing energy functionals with contributions including that from the bulk, computed in the RMF theory \cite{shen} or  Brueckner-Hartree-Fock theory \cite{baldo} for uniform nuclear matter from the neutron and proton densities, surface, exchange (if needed), electron and Coulomb terms.  Microscopic simulations can also provide a deeper knowledge, see \cite{hor,pastanpa}. As a first approach, the present contribution can provide a preliminary view on the species composition in a $\beta-$equilibrated gaseous system using an effective description.  In this spirit we will consider a prescribed minimal coupling of these resonances to mesonic fields and solve the set of self-consistent equations arising. Within this framework, tetraneutrons, being scalar entities with non zero baryon charge, can be represented by a single-component complex field. 

By using an approach based on specifically targeting resonant states \cite{widthQFT} and taking into account the finite quantum mechanical width of these particles one could explicitly introduce this ingredient to the underlying RMF model. In particular and as a result, the calculation of the contribution of tetraneutrons to thermodynamic quantities, such as pressure or energy density, is performed in a different fashion than for other non-resonant states due to the inclusion of a mass distribution function $\rho_{4n}(m)$. This approach has been  applied in various versions of the hadron resonance gas model for the description of experimental hadron multiplicities produced in heavy ion collisions \cite{HRGM1,HRGM2}. Technically, it involves an integration over masses exceeding the dominant decay channel threshold, $m_{4n}^{th}$, while $\rho_{4n}(m)$ gives an integration weight. In this contribution we set $m_{4n}^{th}=4 m_n$, with $m_n$ being the neutron mass, and use a relativistic Breit-Wigner distribution centered at $m=m_{4n}\equiv4m_n+E_{4n}$ with an associated  width $\Gamma_{4n}$ to describe the spread of the tetraneutron mass as we discuss later. 

The structure of this contribution is as follows. In Sec. 2 we present the Lagrangian model used and describe the hadronic content including bosonic and fermionic species.  We introduce the formalism to treat the mass spread of the tetraneutron resonances under the approximation taken.  Later, in Sec. 3, we focus on the effect of the tetraneutron condensate and discuss its impact on the pressure, density and nucleon pairing. We also explain other thermodynamical considerations when solving our set of self-consistent equations. In Sec. 4 we present our results and discuss the baryonic charge fraction carried by the tetraneutron resonance in the range of densities allowing their condensation. We discuss the constraints to bear in mind when considering our results with non-zero fraction of  tetraneutrons, putting especial emphasis in the decay width values that can accommodate such solutions. Finally, in Sec. 5, we give our conclusions.

\section{Lagrangian model and resonances} 

The model Lagrangian considered in our work includes  contribution of nucleons i.e. protons (p) and neutrons (n), electrons (e), light nuclear clusters (deuterons, tritiums, ${}^3He$ nuclei and $\alpha$-particles) as well as tetraneutrons in a zero temperature, electrically neutral and $\beta-$equilibrated medium. Note that the $\Delta(1232)$ resonance will be likely produced  \cite{Delta1,d2} at much higher densities (several times  nuclear saturation density) than those of interest to this work and will not be considered. Thus we will restrict to the exploration of the finite-width effects only regarding  the condensation of tetraneutrons.

In our modelling nuclear interaction is mediated by vector-isoscalar $\omega$, vector-isovector $\rho$ and scalar-isoscalar $\sigma$ mesons. Baryons and stable nuclear clusters are coupled to mesons within a minimal coupling scheme which is exhaustively described in Refs. \cite{RMFclusters1,RMFclusters2,RMFclusters3,RMFclusters4}. Due to the lack of knowledge of the tetraneutron couplings we also choose these resonances to be coupled in the same fashion. In addition, to include further corrections due to finite size of tetraneutrons we consider a spatial radial extent for these resonances up to $R\sim 5$ fm as  it is quite close to their reported size \cite{4n_th11}. This value is significantly larger than $\sim$ 0.4 fm found for nucleons from analysis of hadron multiplicities measured in heavy ion collisions \cite{HRGM1,HRGM2}. Therefore, although we consider the present value of $R$ as an overestimated approximation of the in-medium spatial extent of tetraneutrons, we include it in order to account for effects of their finite size. Coupling strengths are parametrized by constants $g_{j\omega}$, $g_{j\rho}$ and $g_{j\sigma}$  in terms of the nucleon ones ($g_\omega$, $g_\rho$ and $g_\sigma$) as $g_{j\omega}=x_{j\omega}g_\omega$, $g_{j\rho}=x_{j\rho} g_\rho$ and $g_{j\sigma}=x_{j\sigma} g_\sigma$. Let us remind here that these coupling ratios are largely uncertain in our calculation as we will discuss later. Here the index $j$ runs over the species considered. Values of $g_\omega=9.479$, $g_\rho=8.424$ and $g_\sigma=8.487$ are chosen in order to reproduce zero pressure, binding energy per nucleon  equal to 16.3 MeV for the symmetric nuclear matter ground state at saturation density $n_0=0.153$ fm$^{-3}$ as well as the symmetry energy coefficient $a_{sym}=32.5$ MeV. Vacuum rest masses of nucleons and stable nuclear clusters are defined as in Ref. \cite{RMFclusters3} while for mesonic masses we used \cite{MesonMasses}. The mass of tetraneutrons $m_{4n}=4m_n+E_{4n}$ is given in terms of $m_n$ and their excitation energy \cite{4n_exp2}. For electrons we set $m_e=0.511$ MeV. In addition, we assume no magnetic field is present. If that was the case a more detailed analysis including polarization effects \cite{per1,per2} should be included. 

Let us analyze the different contributions in the model Lagrangian used. First, it  includes fermionic terms
\begin{eqnarray}
\label{II}
\mathcal{L}_f=\overline{f}(i\slashed{D}_f-m^*_f)f,
\end{eqnarray} 
of nucleons, electrons, tritiums and $^3He$ nuclei. Here $f$ stands for the Dirac bispinor. Contribution of deuterons is described through the Lorentz vector field $d^{\mu}$ (we sum over repeated indexes)
\small
\begin{equation}
\label{III}
\mathcal{L}_d=\frac{1}{4}(D_{d\mu}d_\nu-D_{d\nu}d_\mu)^*(D_d^\mu d^\nu-D_d^\nu d^\mu)-\frac{1}{2}m^{*2}_d d_\mu^*d^\mu.
\end{equation}
\normalsize
The Lagrangian term for $\alpha$-particles is written as
\begin{eqnarray}
\label{IV}
\mathcal{L}_\alpha=\frac{1}{2}\left[(D_{\alpha\mu}\alpha)^*D^{\alpha\mu}\alpha-m^{*2}_\alpha\alpha^*\alpha\right].
\end{eqnarray}
Finally, the contribution of tetraneutrons is  described with a Lorentz scalar field $\phi$ averaged over the  Breit-Wigner mass distribution 
\begin{equation}
 \rho_{4n}(m)=N\frac{m_{4n}^2\Gamma_{4n}}{(m^2-m_{4n}^2)^2+m_{4n}^2\Gamma_{4n}^2},
\end{equation}
with $N$ a normalization constant. The expression for the resonant tetraneutron Lagrangian is 
\small
\begin{equation}
\mathcal{L}_{4n}=\int\limits_{m^{th}_{4n}}^\infty dm~\frac{\rho_{4n}(m)}{2} \left[D_{4n\mu}^*\phi^*D_{4n}^\mu\phi-(m+\delta m_{4n}-g_{4n\sigma}\sigma)^2\phi^*\phi\right].
\label{V}
\end{equation} 
\normalsize
In our prescription, covariant derivatives are defined as
\begin{eqnarray}
\label{VI}
iD^\mu_j=i\partial^\mu-g_{j\omega}\omega^\mu-g_{j\rho~}\vec{I}_j\cdot\vec{\rho~}^\mu,
\end{eqnarray} 
where $\vec{I}_j$ is the isospin vector of the corresponding particle species. The isospin third component $I^3_j=Q_j-\frac{B_j}{2}$ is defined through the electric $Q_j$ and baryonic $B_j$ charges. In the case of nucleons medium masses are $m^*_j=m_j-g_{j\sigma}\sigma$ while for stable nuclear clusters, instead, $m^*_j=m_j+\delta m_j-g_{j\sigma}\sigma$. In addition, in our setting the medium mass of tetraneutrons is defined as
\begin{eqnarray}
\label{IX}
m^{*2}_{4n}=
\int\limits_{m^{th}_{4n}}^\infty dm~\rho_{4n}(m)~(m+\delta m_{4n}-g_{4n\sigma}\sigma)^2.
\end{eqnarray}
Note, that for $\Gamma_{4n}=0$ the expression of $m^*_{4n}$ recovers the form of stable species since the mass distribution $\rho_{4n}(m)$ acts as Dirac $\delta$-function of argument $m-m_{4n}$.

It is already known that effective medium masses of stable nuclear clusters suffer a shift of their binding and excitation energies from the Pauli blocking corrections \cite{RMFclusters1}. Following the formalism of Ref. \cite{RMFclusters4} we write the corresponding mass shift (including tetraneutrons) as
\begin{eqnarray}
\label{I}
\delta m_k=\frac{Z_k}{n_0}(m_p n_p-\epsilon_p)
+\frac{N_k}{n_0}(m_n n_n-\epsilon_n),
\label{shif}
\end{eqnarray}
where the index $k$ labels the aforementioned species, $Z_k$ and $N_k$ are their proton and neutron numbers, while $n_N$ and $\epsilon_N$ with $N=n,p$ represent the particle and energy densities of gas nucleons of a given sort. It is worth mentioning at this point that this correction is included in \cite{RMFclusters4} for light systems starting from $A=2$ (deuteron) even if it seems to be more justified for medium to heavy nuclei. However, it has proven to give a reasonable effective energy shift for light clusters and, in the same spirit, we introduce this Pauli blocking shift for the tetraneutron resonance. We will nevertheless later discuss the validity of our ansatz. Further, a few-body treatment would involve more refined calculations such as those claimed in \cite{deltuva} where they use a method based on an exact integral version of the Faddeev-Yakubovsky equations governing the 4-fermion system proposed by Alt, Grassberger, and Sandhas (AGS) \cite{ags} or in \cite{beyer} for $\alpha$-particles, however the high complexity of our many-body system prevents its application in this context.

Mesonic contributions to the Lagrangian are the ones of free vector ($\omega$ and $\rho$) and scalar ($\sigma$) fields in addition to the well known self-interaction of $\sigma$-field  \cite{RMFclusters2,RMFclusters3,RMFclusters4,RMFdibaryons2,Delta1,d2,MesonMasses,Qcounting}
\begin{equation}
U_\sigma=-\frac{b}{3}m_n(g_\sigma\sigma)^3-\frac{c}{4}(g_\sigma\sigma)^4.
\end{equation}
We set $b=6.284\times 10^{-3}$ and $c=-3.492\times 10^{-3}$ so the present model fits the incompressibility factor $K_0 = 250$ MeV and the effective nucleon mass $m_N^*=0.75m_N$ at saturation density of symmetric nuclear matter. Note that, although allowed, we do not include other possible non-linear terms involving  $\omega$ and $\rho$ mesons in order to keep our  modelling simple. Regarding the density dependence of the symmetry energy slope we find it to be $L=91.2$ MeV  in our model. This is  somewhat larger than the global average value around $L=61$ MeV and more in line with values arising from studies of nuclear masses \cite{nm} and heavy-ion collisions  \cite{hi}, see Fig. 24 in \cite{dani}.

Isoscalar couplings of stable clusters are set as in Ref. \cite{RMFclusters3}. This set up is consistent with the values of their binding and dissociation energies \cite{RMFclusters1} as well as with experimental predictions of the Mott transition densities \cite{Mott_exp}.

A more careful discussion involves the resonant tetraneutron states. At the moment there is neither theoretical nor experimental information about their coupling strengths. Therefore, the existing degree of uncertainty only permits performing a study in an exploratory fashion. In this spirit and following \cite{RMFclusters4}  we will consider the two sets shown in Table \ref{table2}  where tetraneutron-meson couplings  are set $x_{4n\omega}/A=1$ and $x_{4n\sigma}/A=0.85$  or $1$ with $A=4$.  As we will later discuss the  $x_{4n\sigma}$ parameter is the most influential in our solution and if taken larger than the quoted values it could overpredict tetraneutron onset densities. Isovector couplings are set the same for all particle species, i.e. $x_{j\rho}=1$ except electrons ($x_{e\omega}=x_{e\rho}=x_{e\sigma}=0$). This simple parametrization in terms of fractional ratios $x_{ij}$ is well tested and proven to be successful in many previous works \cite{RMFclusters1,RMFclusters2,RMFclusters3,RMFclusters4,Delta1}. 

\begin{table}
\begin{tabular}{|c|c|c|c|c|c|c|c|c|}
\hline
Set &$x_{4n\sigma}/4$&$x_{4n\omega}/4$&$(n^{os}_{4n}-n^{dis}_{4n})/n_0$&$\Gamma_{4n}$ [MeV]  \\ \hline
 A  &       1.0      &         1.0    &          0.13 - 0.20           &        22.0         \\ \hline
 B  &       0.85     &         1.0    &          0.05 - 0.14           &         9.0         \\ \hline
\end{tabular}
\caption{ Set of different values of $x_{4n\omega}$, $x_{4n\sigma}$ and $\Gamma_{4n}$ used in this work  for both values of $R_{4n}=0,5$ fm considered. Onset and dissolution densities of tetraneutrons $n^{os}_{4n}$, $n^{dis}_{4n}$ are also shown.}
\label{table2}
\end{table}

The spatial extent of tetraneutrons is probably larger than the one of nucleons and bound light nuclear clusters \cite{4n_th11}. This gives rise to additional repulsion between these resonances, which is accounted for by the vector meson driven repulsion and Pauli blocking only in part. In addition, the finite size of tetraneutrons can lead, in principle, to their overlapping. Including an excluded volume correction can effectively prevent this and take into account meaningfully the spatial extent of tetraneutrons as we will see later. The order of magnitude estimate of the baryonic density at which tetraneutrons overlap can be obtained with their typical estimated size $R\sim 5$ fm \cite{4n_th11} and corresponding eigenvolume $V_{4n}=\frac{4\pi}{3}R^3$. Using a typical fraction of baryonic  charge carried by tetraneutrons $4n_{4n}/n_B$ about 0.1 (see Fig. 3) and $n_{4n}=V_{4n}^{-1}$ we obtain $n_B\simeq\frac{30}{\pi R^3}\simeq 0.08 ~\rm fm^{-3}$, which is well above the range of densities where tetraneutron may condense as we actually find in this work. Thus since these resonances are expected to exist only at small densities, the Van der Waals approximation results sufficiently accurate. A corresponding correction can be introduced to the present model within the prescription of \cite{VdW}. According to it, the effective chemical potential of tetraneutrons is reduced by $4V_{4n}p$, where $p$ is the total pressure.

\section{Effect of Bose condensation of Tetraneutrons}

In our model, $p$ can be written as a sum of partial pressures of non-interacting quasiparticles, with the medium masses $m^*_j$ and effective chemical potentials defined through baryonic $\mu_B$ and electric $\mu_Q$ chemical potentials, mean values of the scalar field $\sigma$ and temporal components of the $\omega$ and $\rho$ the vector fields. In the case of tetraneutrons 
\begin{eqnarray}
\label{IIIa}
\mu^*_{4n}=\mu_B B_{4n}+\mu_Q Q_{4n}-g_{4n\omega}\omega-g_{4n\rho} I^3_{4n}\rho- 4V_{4n}p\,
\end{eqnarray}
while for other pointlike species
\begin{eqnarray}
\label{III}
\mu^*_j=\mu_B B_j+\mu_Q Q_j-g_{j\omega}\omega-g_{j\rho} I^3_j\rho\,.
\end{eqnarray}
An excluded volume of tetraneutrons in Eq. (\ref{IIIa}) is introduced within the Van der Waals approximation. In the grand canonical ensemble this framework leads to the pressure $p_{VdW}(\mu)=p_0(\mu-4Vp_{VdW})$ defined through the eigenvolume of particles $V$ and the pointlike pressure $p_0(\mu)$. Formally, the excluded volume of tetraneutrons appears due to the repulsion between the hard nuclear cores. Therefore, a corresponding negative correction $- 4V_{4n}p$ arises as for any repulsive term. 
Besides, physical chemical potentials are $\mu_j=\mu_B B_j+\mu_Q Q_j$. This form of $\mu_j$ automatically insures $\beta$-equilibrium since in this case $\mu_n-\mu_p=\mu_e$ by construction. Values of $\mu_B$ and $\mu_Q$ are defined jointly by the value of the baryonic density $n_B$ and the requirement of electrical neutrality. Finally, the expression for the total pressure at zero temperature can be written in the form
\begin{eqnarray}
p&=&\sum_f\frac{g_f}{6\pi^2}\int_0^{k_f}\frac{dk~k^4}{\sqrt{m^{*2}_f+k^2}}
+\sum_b\zeta^2_b(\mu^{*2}_b-m^{*2}_b)
\nonumber\\
\label{IV}
&+&\frac{m^2_\omega\omega^2}{2}+\frac{m^2_\rho\rho^2}{2}
-\frac{m^2_\sigma\sigma^2}{2}+U_\sigma,
\end{eqnarray}
where $d_f=2$ is the spin degeneracy factor and the Fermi momentum  $k_f=\sqrt{\mu^{*2}_f-m ^{*2}_f}$.  Note that formally Eq. (\ref{IV}) is an equation to be solved self-consistently in order to find the pressure since it enters on the righthand side indirectly through $\mu^*_{4n}$. The first sum runs over all fermionic species. The second one accounts for deuterons, $\alpha$-particles and tetraneutrons which at zero temperature exist as the Bose-Einstein condensate (BEC) \cite{Kapusta}. Note that this BEC is significantly delocalized since it exists in the lowest energy state and its wave function is characterized by a spatial spread that is much larger than the typical cluster size of a few fm. The real numbers $\zeta_b$ represent amplitudes of zero modes in the field operators of these bosons. Physical values of $\zeta_b$ are obtained by maximization of pressure, i.e. from the condition
\begin{eqnarray}
\frac{\partial p}{\partial \zeta_b}=2\zeta_b(\mu^{*2}_b-m^{*2}_b)=0.
\end{eqnarray}
This yields  either $\zeta_b=0$ or
\begin{eqnarray}
\label{V}
\mu^*_b=m^*_b\,.
\end{eqnarray}
Thus, the BEC contribution to Eq. (\ref{IV}) is always zero. At the same time the density of condensed bosons
\begin{eqnarray}
n_b=\frac{\partial p}{\partial\mu_b}=
\frac{2\zeta^2_b\mu^*_b}{1+4V_{4n}\cdot2\zeta^2_{4n}\mu^*_{4n}}
\end{eqnarray}
has a finite value if the previous equality is fulfilled. Therefore, Eq. (\ref{V}) is a condition for the BEC existence. Note, that the denominator of this expression is caused by the fact that part of the system volume is excluded by non pointlike tetraneutrons. Note that the same denominator appears in expressions for densities of the rest of all species. For fermions it reads
\begin{eqnarray}
n_f=\frac{\frac{g_f}{6\pi^2}\int_0^{k_f} dk~k^3}{1+4V_{4n}\cdot2\zeta_{4n}^2\mu_{4n}^*}.
\end{eqnarray}
In the case of tetraneutrons it ensures that they never overlap. Indeed, even at the maximal packing of  tetraneutrons reached at $\zeta_{4n}^2\mu_{4n}^*\rightarrow\infty$ each of them occupies a cell of volume $V_{\rm cell}=n^{-1}_{4n}=4V_{4n}$, which is four times larger than their eigenvolume.  

The energy density can be obtained from the previous expressions considering contributions of both bosonic and fermionic species, $\varepsilon=\sum_{f,b}\mu_jn_j-p$. Mean mesonic fields can be self-consistently found from conditions of maximal pressure $\frac{\partial p}{\partial\omega}=0$, $\frac{\partial p}{\partial\rho}=0$ and $\frac{\partial p}{\partial\sigma}=0$. At given baryonic density they define all termodynamic quantities of electrically neutral nuclear matter at $\beta$-equilibrium.

%
\begin{figure}[th]
\includegraphics[width=0.85\columnwidth]{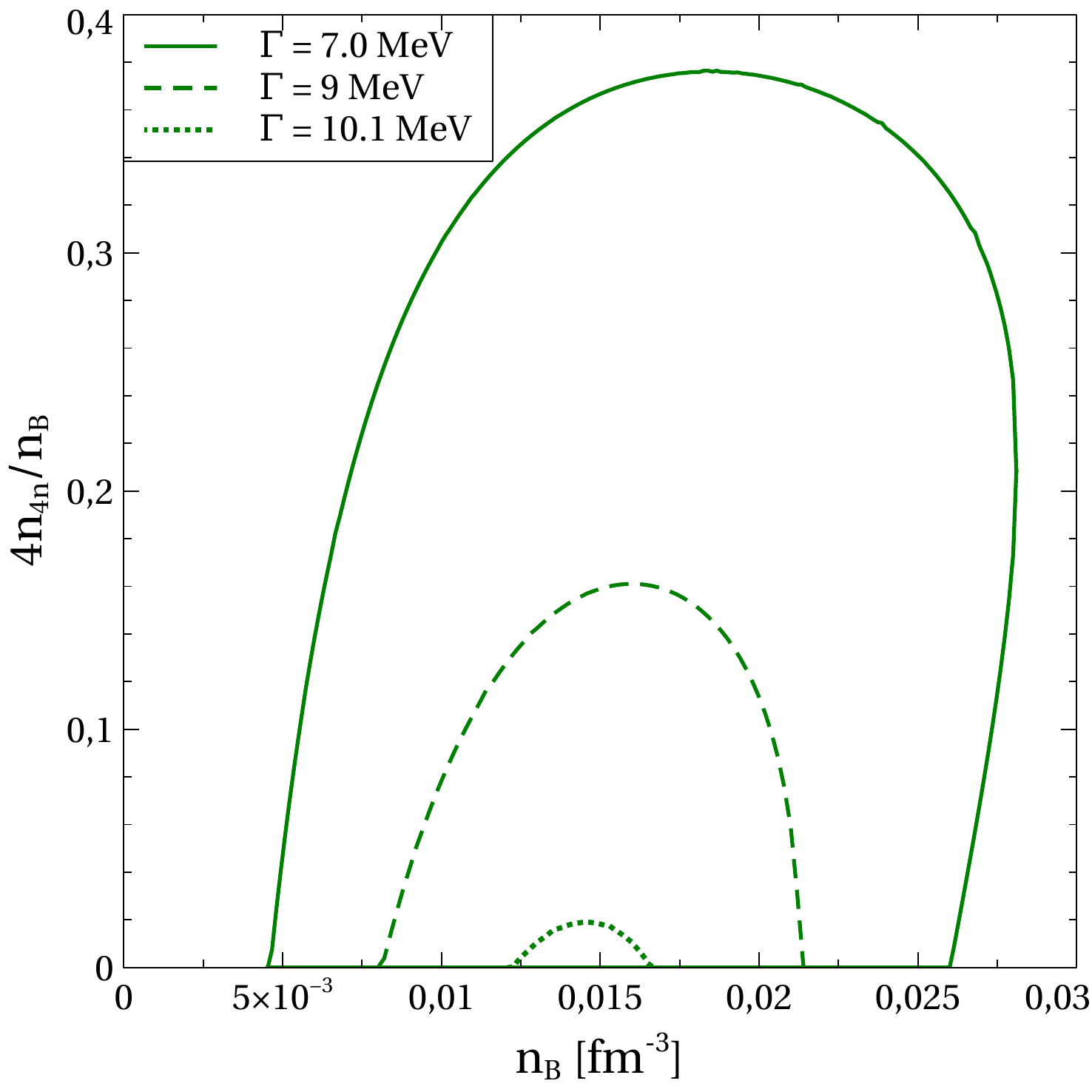}
\caption{ Baryonic fraction of the tetraneutron condensate as a function of baryonic density for three values of $\Gamma_{4n}$ for pointlike ${}^4n$ ($R=0$) and set B. Dashed and dotted lines correspond to physical condensation while solid line depicts the case where $\Gamma_{4n}$ lies outside the physical condensation interval. For this case $\Gamma_{\rm max}=10.2$ MeV and $\Gamma_{\rm min}=8.2$ MeV .}
\label{fig1}
\end{figure}

We paid special attention to the analysis of the possibility of the tetraneutron BEC existence. As mentioned before, there is a large uncertainty regarding $\Gamma_{4n}$ values. Therefore and in order to be practical we first set this parameter equal to an average value $\Gamma_{4n}\simeq 9$ MeV which is close to the inverse vacuum lifetime of the tetraneutron $\tau\sim 10^{-22}$ s as   reported in Ref. \cite{4n_exp2}. However, when considering the isospin asymmetric nuclear medium, $\tau$ could be, in principle, modified by the interaction with neutrons \cite{oset} whereas rare collisions with other particles could be neglected. Thus $\tau$ and, accordingly, $\Gamma_{4n}$ are expected to suffer a possible in-medium widening with typical values belonging to the interval of $\Gamma_{4n}\simeq 10 - 30$ MeV at large densities. In our case, however, the reported densities for condensation are well below nuclear saturation density, see below.

On the other hand, the fraction of baryonic charge carried by the tetraneutron BEC crucially depends on the tetraneutron width. We found that for any realistic value of $x_{4n\sigma}/4\le1$ tetraneutrons can exist only in some limited range of densities. At the same time, for any set of coupling constants and value of the hard-core radius, the  topology of the solution can be different regarding the value of $\Gamma_{4n}$. For example, Fig.  \ref{fig1} shows this fraction as a function of baryonic density calculated for set B i.e. $x_{4n\sigma}/4=0.85$, $x_{4n\omega}/4=x_{4n\rho}=1$ and $R_{4n}=0$ considering  several values of the tetraneutron width. As it is seen from the figure, at small $\Gamma_{4n}$ (solid curve) some baryonic densities  around $n_B\sim 0.0275$ $\rm fm^{-3}$ support the simultaneous existence of two different values of the tetraneutron fractions, which is unphysical. This situation happens only for the tetraneutron widths smaller than some critical value $\Gamma_{\rm min}$.

\begin{figure}[th]
\includegraphics[width=0.85\columnwidth]{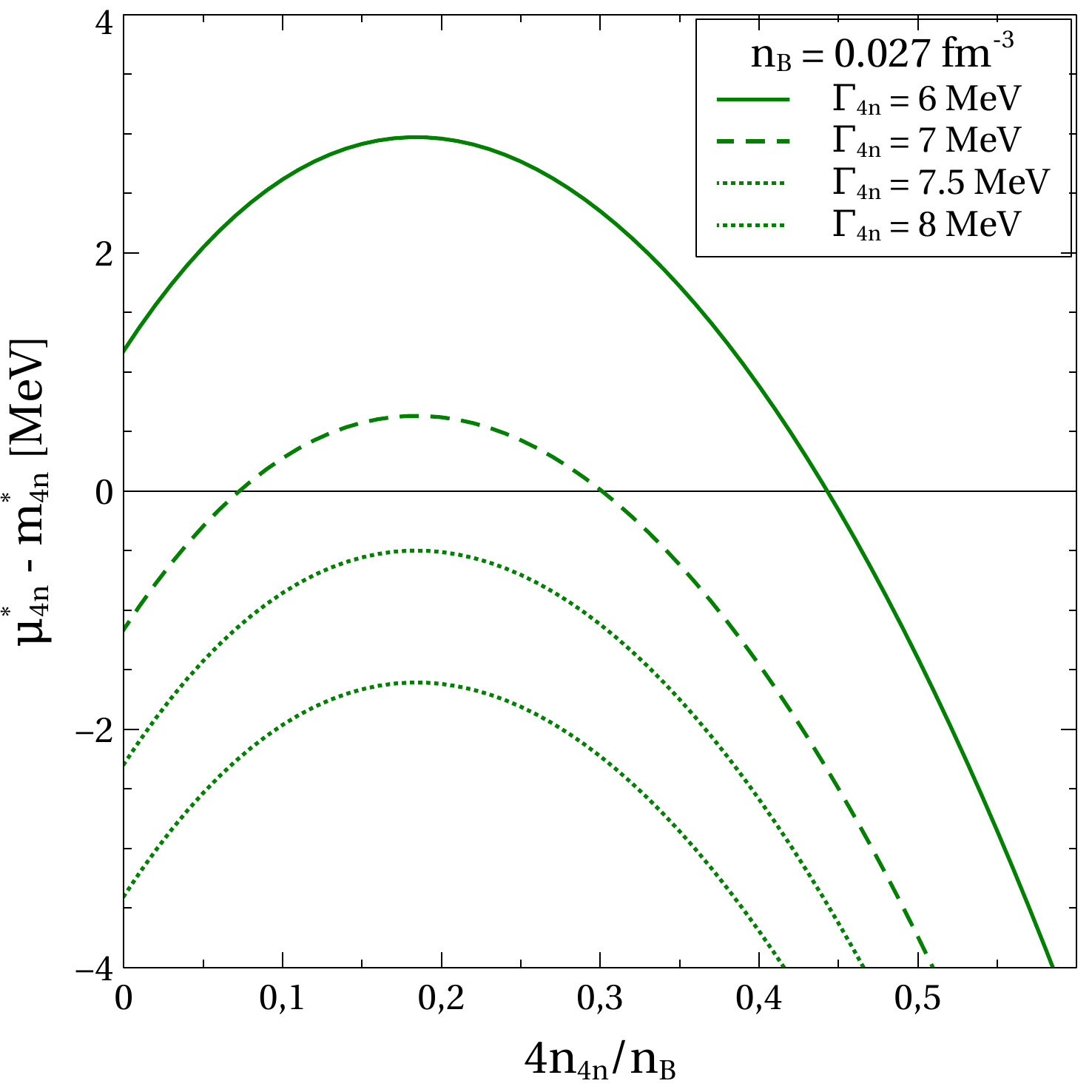}
\caption{ Difference of the effective chemical potential and mass of pointlike tetraneutrons ($R=0$) as function of the baryonic charge fraction carried by them for four values of $\Gamma_{4n}$  and set B. Baryonic density is set $n_B = 0.027~{\rm fm}^{-3}$.}
\label{fig61}
\end{figure}

In order to understand this behaviour let us discuss now the possible appearance of two solutions for a given density of tetraneutrons. The problem of finding $n_{4n}$ can be solved in two steps. The first one corresponds to solving the field equations for $\sigma$, $\omega$ and $\rho$ for some arbitrary amplitude of the zero mode of tetraneutrons $\zeta_{4n}$. With these mean mesonic fields one can find the effective mass $m_{4n}^*$ and chemical potential $\mu_{4n}^*$ of tetraneutrons. Since the amplitude of zero mode and density of tetraneutrons are related by Eq. (15), then $\mu_{4n}^*$ and $m_{4n}^*$ can be considered functions of $n_{4n}$. On the second step $n_{4n}$ can be found by requiring $\mu_{4n}^*=m_{4n}^*$, which agrees with the condition (14) for the existence of the tetraneutron BEC. Fig. \ref{fig61} illustrates the situation when at small $\Gamma_{4n}$ the condition $\mu_{4n}^*=m_{4n}^*$ can be fulfilled in different manners. This shows how the two solutions for a condensate of tetraneutrons with $\Gamma_{4n}=7$ MeV at $n_B = 0.027~{\rm fm}^{-3}$ depicted in Fig. 1 arise. At the same time, the larger values of the tetraneutron width suppress these aggregates making their condensation impossible since in this case the condition $\mu_{4n}^*=m_{4n}^*$ can not be fulfilled by any value of $n_{4n}$.

%

\begin{figure}[!]
\includegraphics[width=0.85\columnwidth]{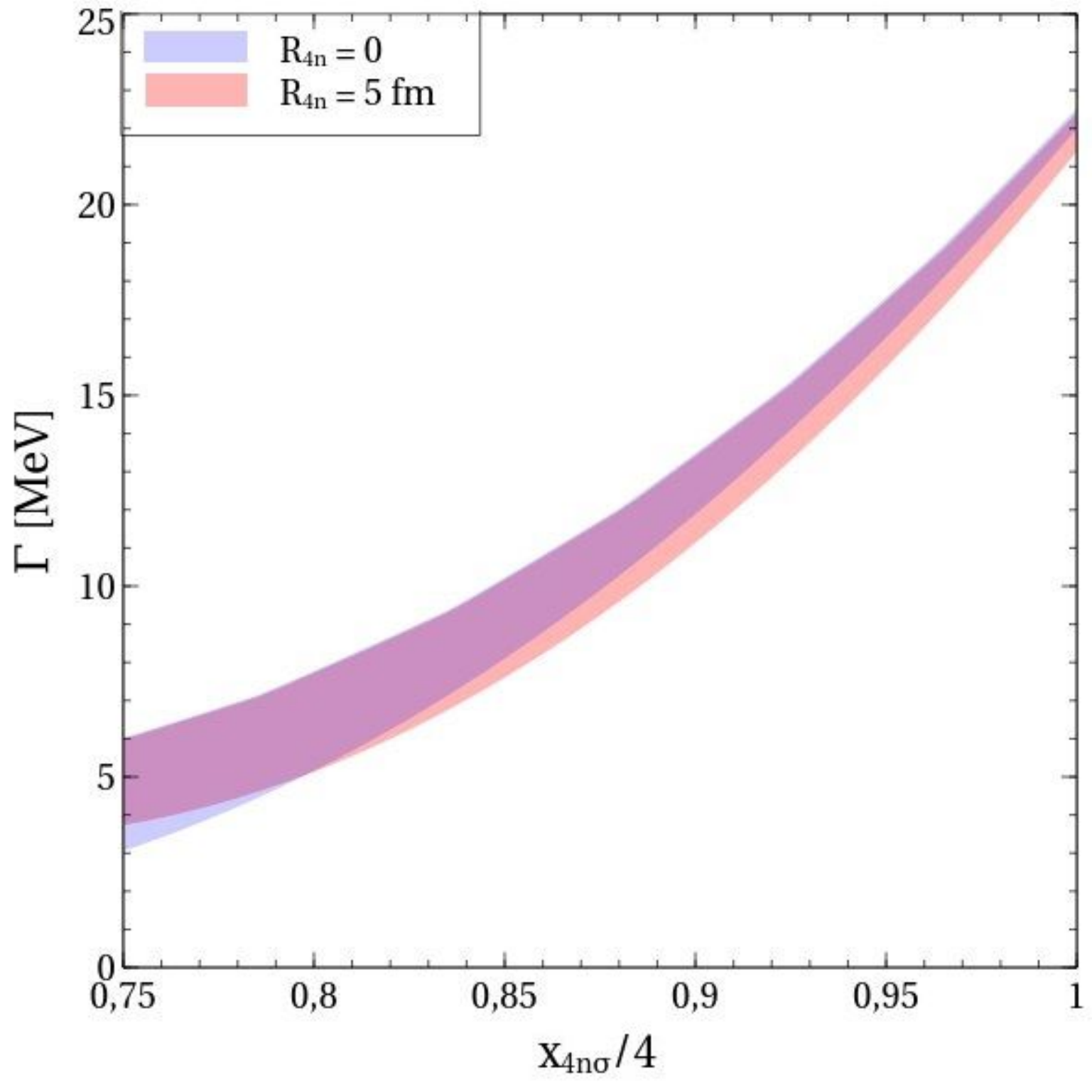}
\caption{Colour bands depicting the allowed regions for pointlike ($R=0$) and finite-size ($R=5$ fm) tetraneutron condensate as a function of $x_{4n\sigma}/4$ at $x_{4n\omega}/4=x_{4n\rho}=1$. Overlapping regions appear in magenta.}
\label{fig11}
\end{figure}

 Therefore, we conclude that a physically meaningful $\Gamma_{4n}$ should be larger than $\Gamma_{\rm min}$. Furthermore, in Fig.  \ref{fig1}  large values of $\Gamma_{4n}$ (dotted curve) make the fraction of tetraneutrons tiny, while at $\Gamma_{4n}$ exceeding some critical value $\Gamma_{\rm max}$, they totally disappear. This allows us to conclude that for a single and physically correct solution with non zero fraction of tetraneutrons (dashed and dotted curves  on Fig. \ref{fig1}) we must have $\Gamma_{\rm min}\le\Gamma_{4n}\le\Gamma_{\rm max}$. We show in Fig. \ref{fig11} the coloured blue (pink) bands of $\Gamma_{4n}$ providing existence of a condensate of pointlike (finite size) tetraneutrons as a function of $x_{4n\sigma}/4$ at $x_{4n\omega}/4=1$. This phenomenological constraint indicates the need for a careful determination of the in-medium width $\Gamma_{4n}$ and its dependence on the coupling $x_{4n\sigma}$.

\section{Results} 

We now comment on the results found in our calculation. Let us first remind that, in what follows, our analysis and calculations are performed for the two selected values of coupling parameter sets, A and B, and two values of spatial tetraneutron extent, $R$, and decay width, $\Gamma_{4n}$  \cite{natu}. Since set A predicts larger baryonic densities where these resonances may exist, increased $\Gamma_{4n}$ values are allowed (see Fig. \ref{fig1}).  It is important to notice that in what follows we have chosen values of $\Gamma_{4n}$ allowing a physical BEC for both sets A and B but if values are outside the interval, tetraneutrons would not be able to exist as predicted in our scenario. Values of  $x_{4n\omega}$ and $x_{4n\sigma}$ which correspond to different onset,   $n^{\rm os}_{4n}$, and dissolution densities, $n^{\rm dis}_{4n}$, of tetraneutrons are listed in Table \ref{table2}.

\begin{figure}[!]
\includegraphics[width=0.85\columnwidth]{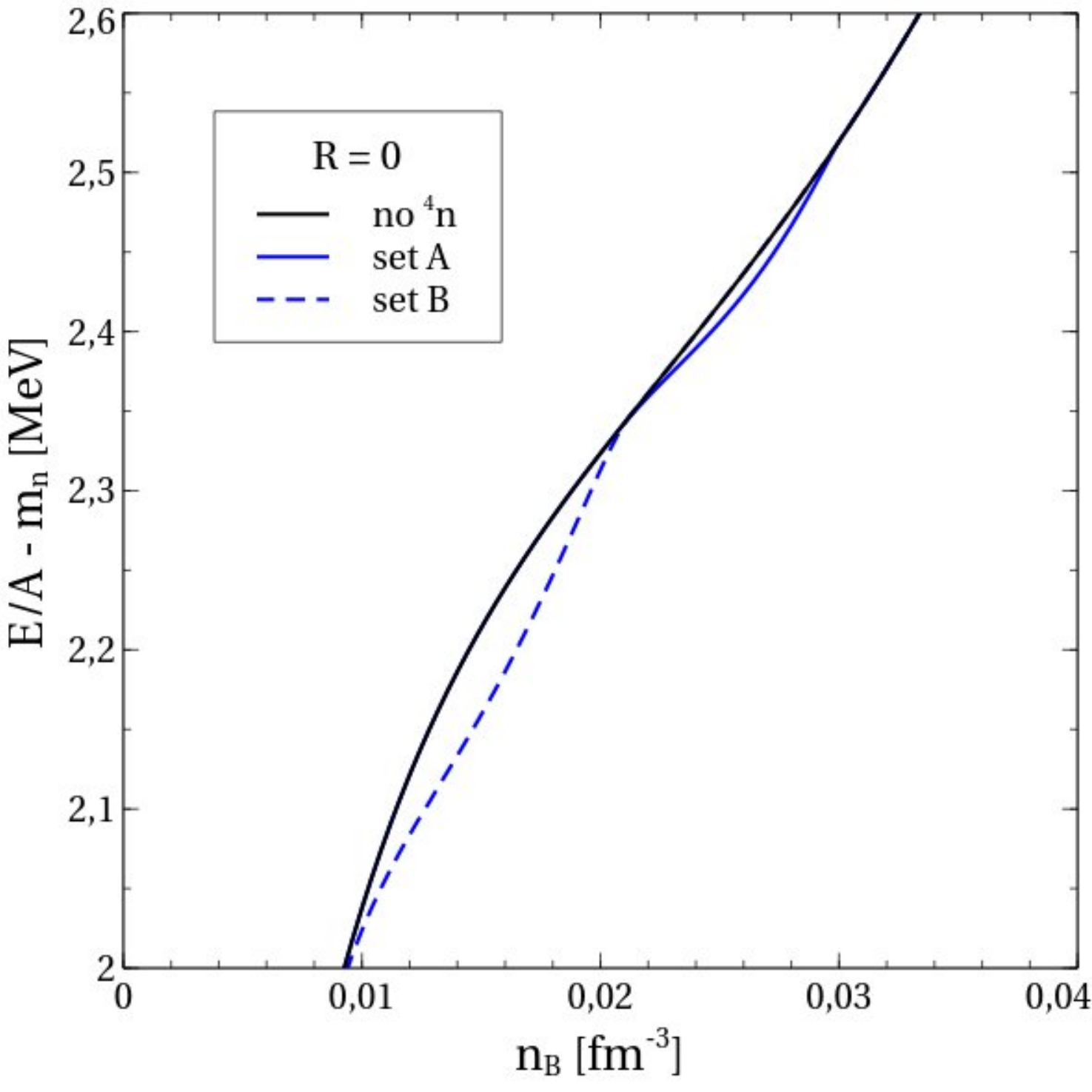}
\caption{ Energy per nucleon as a function of baryonic density for parameter sets A and B and tetraneutron free matter. Tetraneutrons are included in a  pointlike ($R=0$) approximation. The density range corresponding to the tetraneutron condensation signals the more energetically stable states. }
\label{fig31}
\end{figure}

In Fig. 4 we show the total energy per nucleon $\epsilon/n_B-m_n$ for set A (solid line) and set B (dashed line) as a function of baryonic density. We also depict the energy of a tetraneutron free system for reference. A pointlike treatment for tetraneutrons has been used. It can be clearly seen that for the two parameter sets used, matter with tetraneutrons is energetically favoured over that where tetraneutrons are not allowed. In other words, the presence of tetraneutrons  in neutron rich matter is energetically favoured even despite the short lifetime of these resonances.

In order to better understand the role of tetraneutrons in the nuclear system under study we plot in Fig. 5 the
difference of effective chemical potentials and masses of nuclear clusters in the case of pointlike ($R = 0$) tetraneutrons as a function of baryonic density for sets A and B. It is convenient to analyse this quantity since a nuclear species $j$ can exist only if $\mu_j^*-m_j^*\ge0$ in the case of fermions or $\mu_j^*-m_j^*=0$ in the case of bosons. The region of the tetraneutron BEC existence is defined by the condition $\mu_{4n}^*-m_{4n}^*=0$ appearing as horizontal line segments of dashed (set A) and solid (set B) green curves. It is clearly seen that in the case of set A all nuclear clusters have $\mu_j^*-m_j^*<0$ in the region of the tetraneutron BEC. In other words, tetraneutrons do not coexist with these clusters. The same situation happens in the case of set B for all clusters except $\alpha$-particles, which can exist simultaneously with tetraneutrons in a narrow range of $n_B=0.007 - 0.01~{\rm fm}^{-3}$. At the same time, fractions of $\alpha$ and ${}^4n$ in the overlap region are so small that their impact on each other is almost absent. This explains why for deuterons, tritiums, ${}^3$He and $\alpha$-particles $\mu_j^*-m_j^*$ is the same for sets A and B. As we have verified, accounting for the finite size of tetraneutrons leads to the decoupling of regions where stable nuclear clusters and tetraneutrons can coexist. This happens due to the increase of the tetraneutron onset density. As a generic conclusion we can say that the present couplings of stable nuclear clusters disfavour their coexistence with tetraneutrons.


\begin{figure}[th]
\includegraphics[width=0.85\columnwidth]{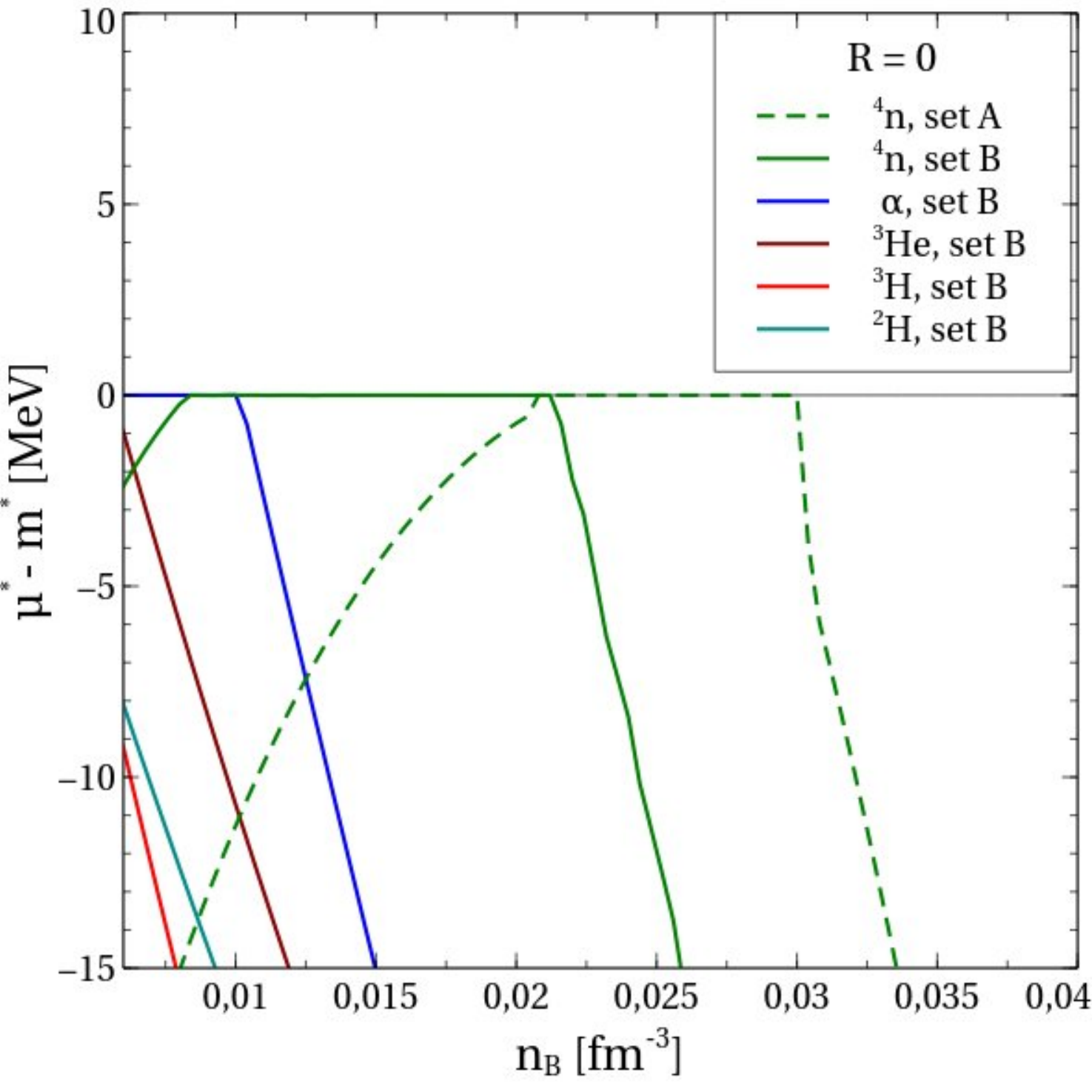}
\caption{
Difference of effective chemical potentials and masses of different species in the case of pointlike ($R = 0$) tetraneutrons as a function of baryonic density for sets A and B. The regions of the tetraneutron BEC is shown by the horizontal line sigments of dashed (set A) and solid (set B) green curves signaling the vanishing value for the corresponding density range. Note, that ${}^3$He, deuterium and tritium do not condensate nor coexist with tetraneutrons, while for $\alpha$-particles in the case of set B there is an narrow overlapping region at $n_B=0.007 - 0.01~{\rm fm}^{-3}$.
}
\label{fig41}
\end{figure}

\begin{figure}[th]
\includegraphics[width=0.85\columnwidth]{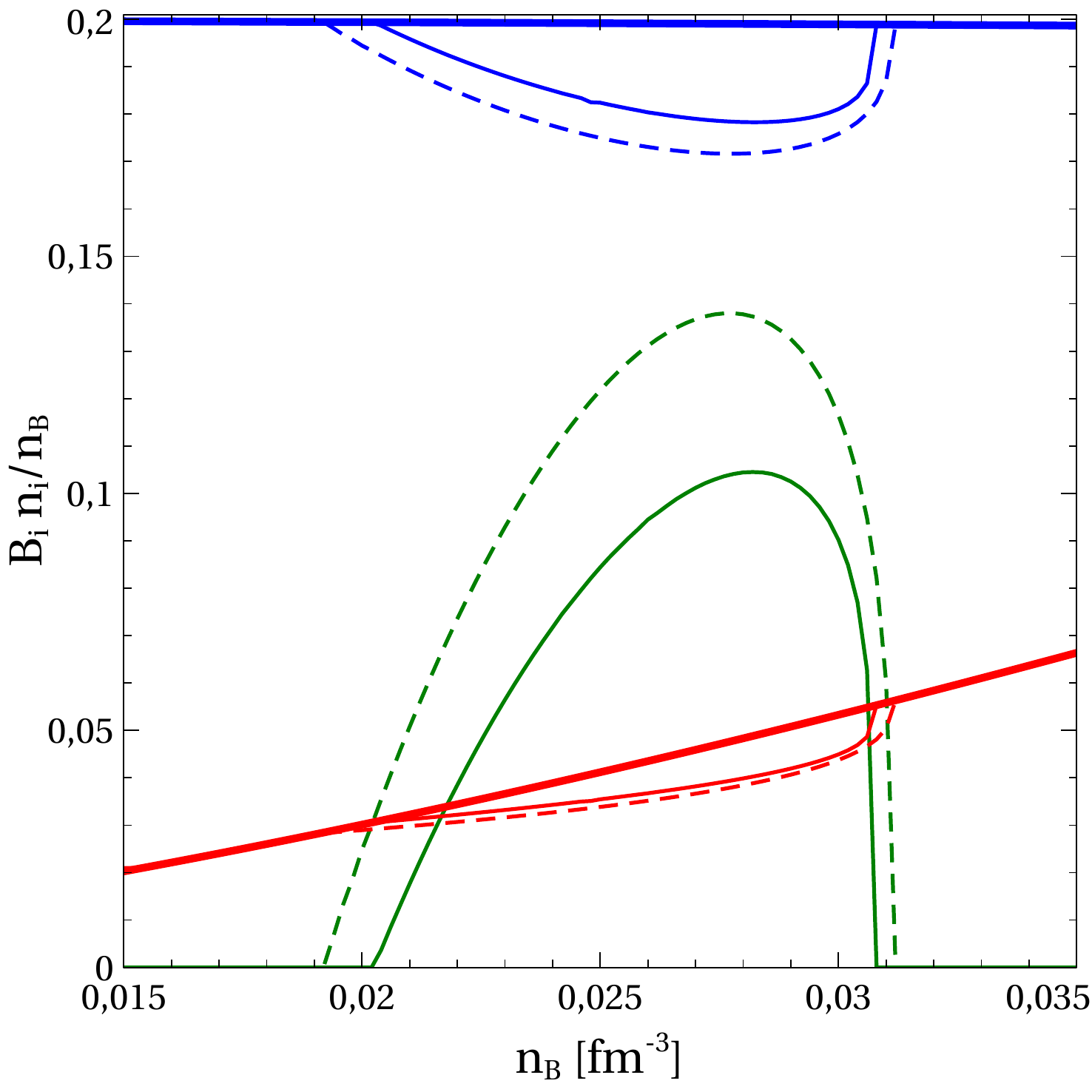}
\includegraphics[width=0.85\columnwidth]{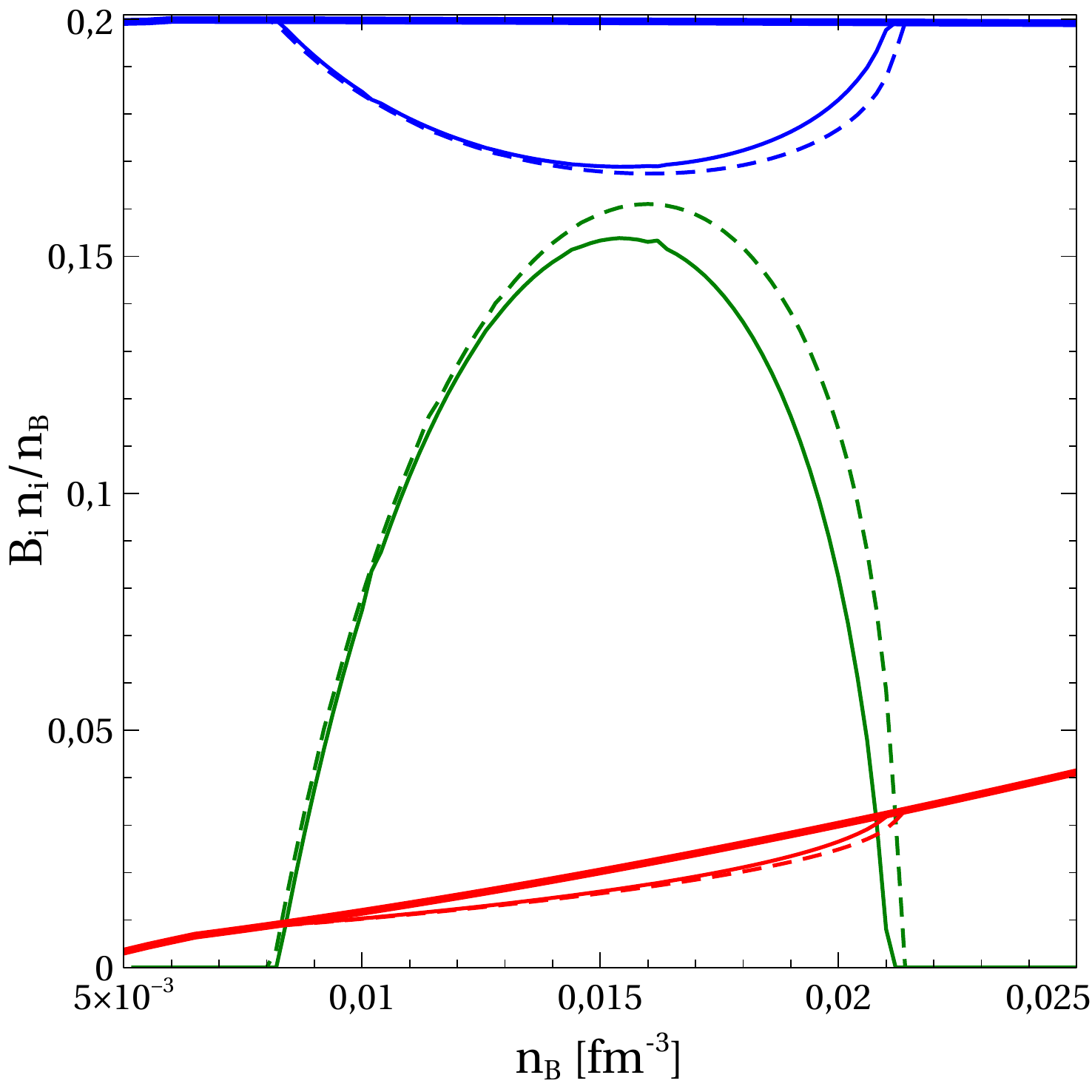}
\caption{  Baryonic charge fraction for the n (blue), p (red) and tetraneutron (green) components as functions of baryon density for set A (upper panel) and  B (lower panel). For each species we plot $R=0$ (dashed line) and $R=5$ fm (solid line) cases along with the ${}^4n$ free solution (thick solid). The rest of species are omitted since they mostly do not coexist with tetraneutrons (see Fig.5). We have scaled n curves by a factor $1/5$ and p curves by a factor of $10$ in order to facilitate the reading. See text for details.}
\label{fig12}
\end{figure}

In Fig. \ref{fig12} we show the baryonic charge fraction  for set A (upper panel) and set B (lower panel). We depict the different components, i.e.  n (blue), p (red) and tetraneutron (green) species  as a function of baryon density $n_B$ calculated for $R=0$ (dashed line) and $R=5$ fm (solid line).  We also include the ${}^4n$ free solution (thick solid line). We have scaled n curves by a factor $1/5$ and p curves by a factor of $10$ in order to facilitate the reading. In our $\beta-$equilibrated system the fraction of protons remains tiny at all densities and the appearance of light clusters is suppressed due to the combined effect of the negative contribution of electric  chemical potential and the selected set of coupling parameters. We can see that onset (dissolution) densities of tetraneutrons are larger (smaller) when including  finite volume corrections.

Note that the particle density fraction of ${}^4n$ is obtained dividing each value on the tetraneutron curves by $B_{4n}=4$ since their particle number density is $B_{4n}$ times smaller than the corresponding baryonic charge  density. We can see that for the two parameter sets used in this work $x_{4n\sigma}/4\lesssim 1$ and  ${}^4n$ are restricted to the lower densites in a small fraction up to $\sim 4\%$ of the baryon density. In the case of the depicted set A (upper panel) ${}^4n$  exist in the range $n_B\simeq (0.02-0.031)\rm \, fm^{-3}$,  in line with typical Mott densities. For densities larger than those it is not so energetically favourable to gather charge into these resonances due to Pauli blocking. 
\begin{figure}[th]
\includegraphics[width=0.85\columnwidth]{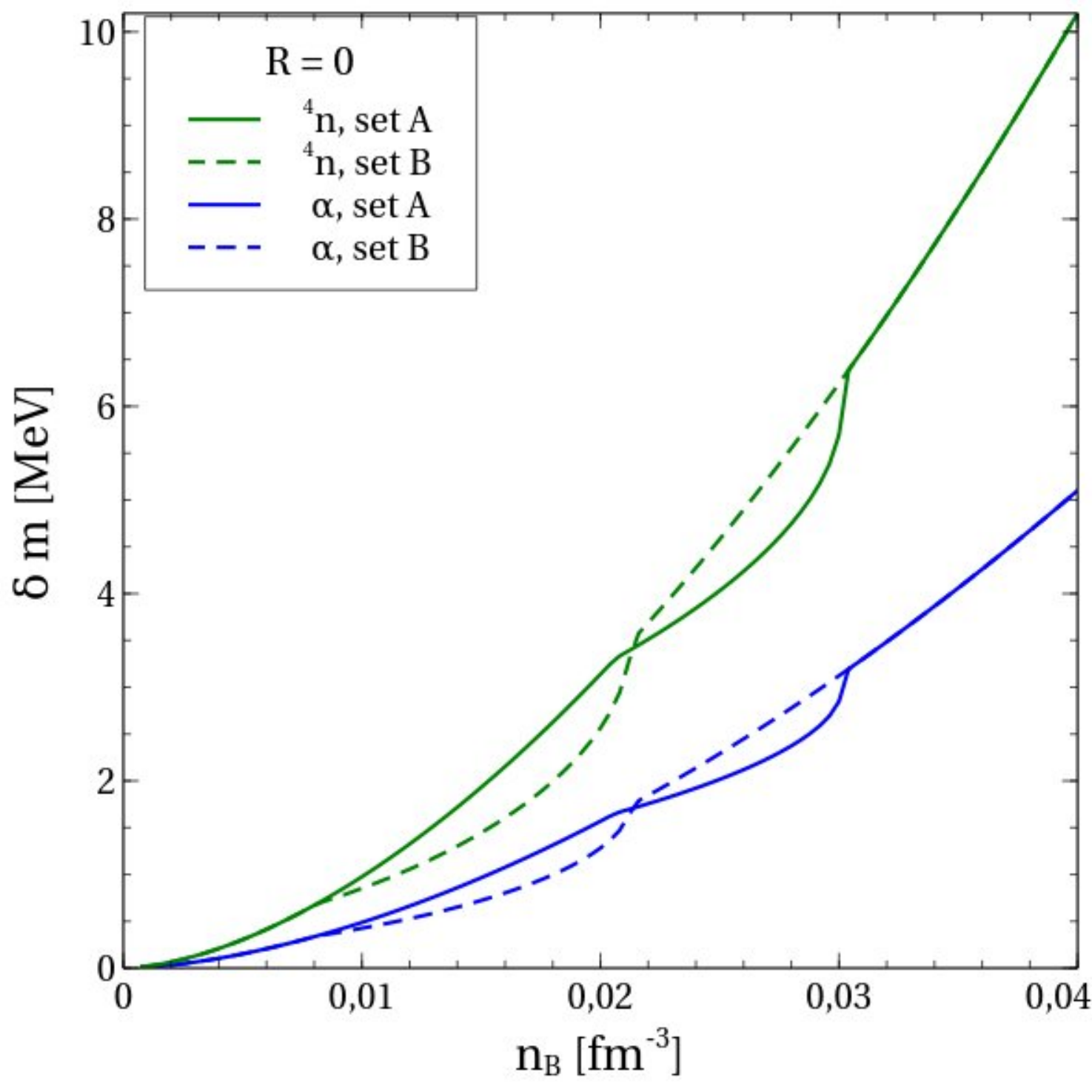}
\caption{Mass shift for pointlike ($R=0$) tetraneutron (green) and alpha particle (blue) species for parameter sets A (solid line) and B (dashed line) as a  function of baryonic density $n_B$.}
\label{fig21}
\end{figure}

\begin{figure}[th]
\includegraphics[width=0.85\columnwidth]{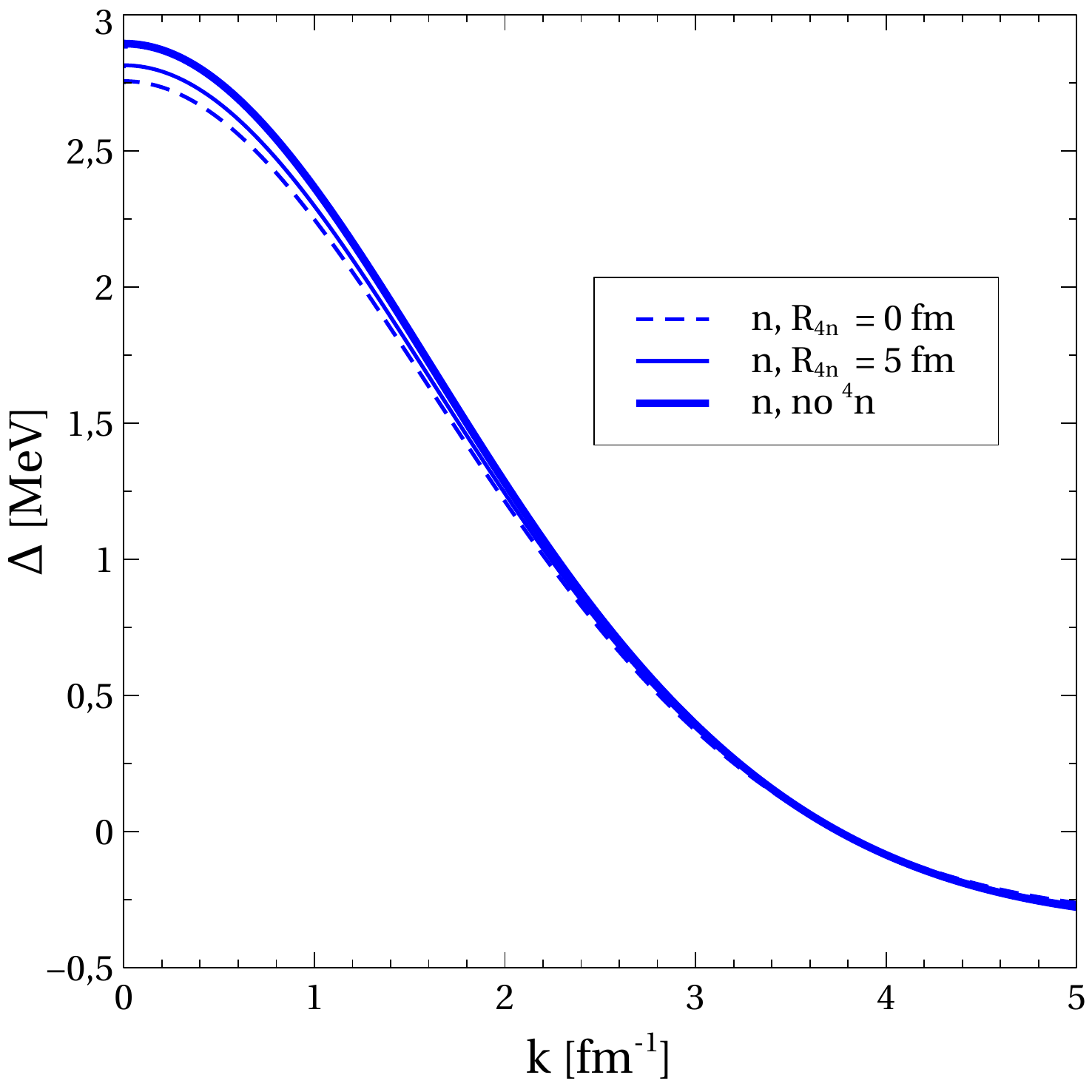}
\includegraphics[width=0.85\columnwidth]{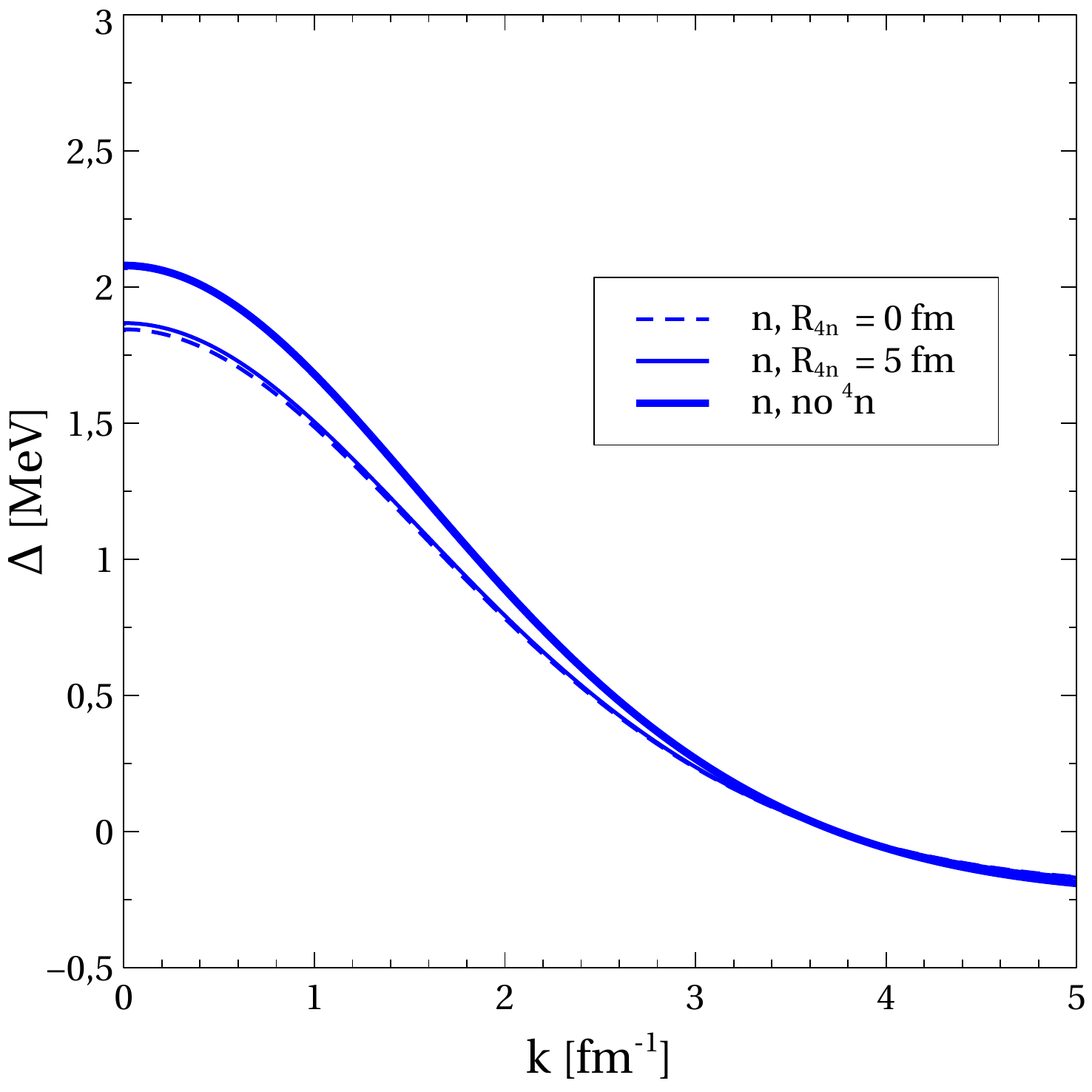}
\caption{Neutron pairing gap as a function of momentum for sets A (upper panel) and B (lower panel). We depict the cases of pointlike tetraneutrons  (thin dashed  line) and $R=5$ fm (thin solid line). The case with no tetraneutrons is shown with thick solid line.  For the tetraneutron free case neutron Fermi momenta $k_F = 0.9323$ $\rm fm^{-1}$ (set A) and $k_F = 0.7625$ $\rm fm^{-1}$ (set B) correspond to $n_B = 0.0275$ fm$^{-3}$ and $n_B = 0.0150$ fm$^{-3}$, respectively. These later values lie close to the maximum of the baryonic charge fraction for tetraneutrons depicted in Fig. \ref{fig12}.}
\label{fig2}
\end{figure}

To illustrate this we show in Fig. 7  the mass shift induced by Pauli blocking for the case of two sets of couplings, A and B, for pointlike tetraneutrons and for $\alpha$-particles. The effect of this term in Eq. (8) clearly induces an extra energetic cost for composite species since $\delta m>0$. It is also seen that in the case of tetraneutrons it is about two times stronger than for the case of $\alpha$-particles. This is caused by the fact that most of the Pauli blocking induced shift of mass comes through neutrons, while the contribution of protons can be neglected due to their small density. Consequently, with a good accuracy we can conclude that $\frac{\delta m_{4n}}{\delta m_\alpha}\simeq\frac{N_{4n}}{N_{\alpha}}=2$. However, as we will see later there are additional dependencies on the energy density that can overcome this fact and lead to an energetically favourable solution where tetraneutrons are indeed present.


In order to further explore the microscopic consequences of the presence of tetraneutrons we have studied the nucleon pairing into spin-zero Cooper pairs. We have selected the most attractive ${}^1S_0$ channel using for this purpose the strategy of Ref. \cite{RMFpairing}. Note that we consider the BCS approximation although more refined treatments are indeed possible \cite{sedrakian,lombardo} quoting in particular those including short-range and long-range correlations to account for medium effects and polarization  \cite{dick}. The dependence of the nucleon pairing gap $\Delta_N$ on its momentum $k$ is defined by the gap equation
\footnotesize
\begin{eqnarray}
\label{XVI}
\Delta_N(k)&=&-\frac{1}{\pi}\int\limits_0^\infty dk'
\frac{{k'}^2V(k,k')\Delta_N(k')}
{\sqrt{(\epsilon_N(k',m^*_N)-\mu^*_N)^2+\Delta^2_N(k')}},
\end{eqnarray}
\normalsize%

where $\epsilon_N(k',m^*_N)=\sqrt{k'^2+m^{*2}_N}$ and the matrix elements of the two-nucleon interaction potential $V$ are 
\begin{eqnarray}
\label{XVII}
V(k,k')&=&\int\limits_0^\infty dr~r^2 j_0(kr)V(r)j_0(k'r).
\end{eqnarray}
Here $j_0$ denotes the first kind Bessel function of order zero. The consistency with the present model Lagrangian is provided by the Yukawa parametrization of $V$, which includes repulsive contributions from the $\omega$ and $\rho$ mesons as well as an attractive one from the $\sigma$ meson. Thus, in the coordinate space
\begin{equation}
\label{XVIII}
V(r)=\frac{A}{r}\left[g_\omega^2e^{-m_\omega r}+
\left(\frac{g_\rho}{2}\right)^2e^{-m_\rho r}-g_\sigma^2e^{-m_\sigma r}\right].
\end{equation}
A phenomenological parameter $A=0.0435$ is chosen in order to provide reasonable characteristics of this potential. Its minimum is located at $r= 0.61$ fm and has a depth of 50 MeV. The factor $\frac{1}{2}$ in the $\rho$-meson term comes from the nucleon isospin. Note, that such a potential can be derived within the one boson exchange (OBE) approximation \cite{PotentialOBE}. Safely, as it has been shown in \cite{jensen}  for standard BCS calculations, gap energies are quite similar for different realistic interactions and we choose this parametrization for the sake of simplicity. We believe this treatment captures the essence of the nucleon pairing in our diverse population scenario.

\begin{figure}[th]
\includegraphics[width=0.85\columnwidth]{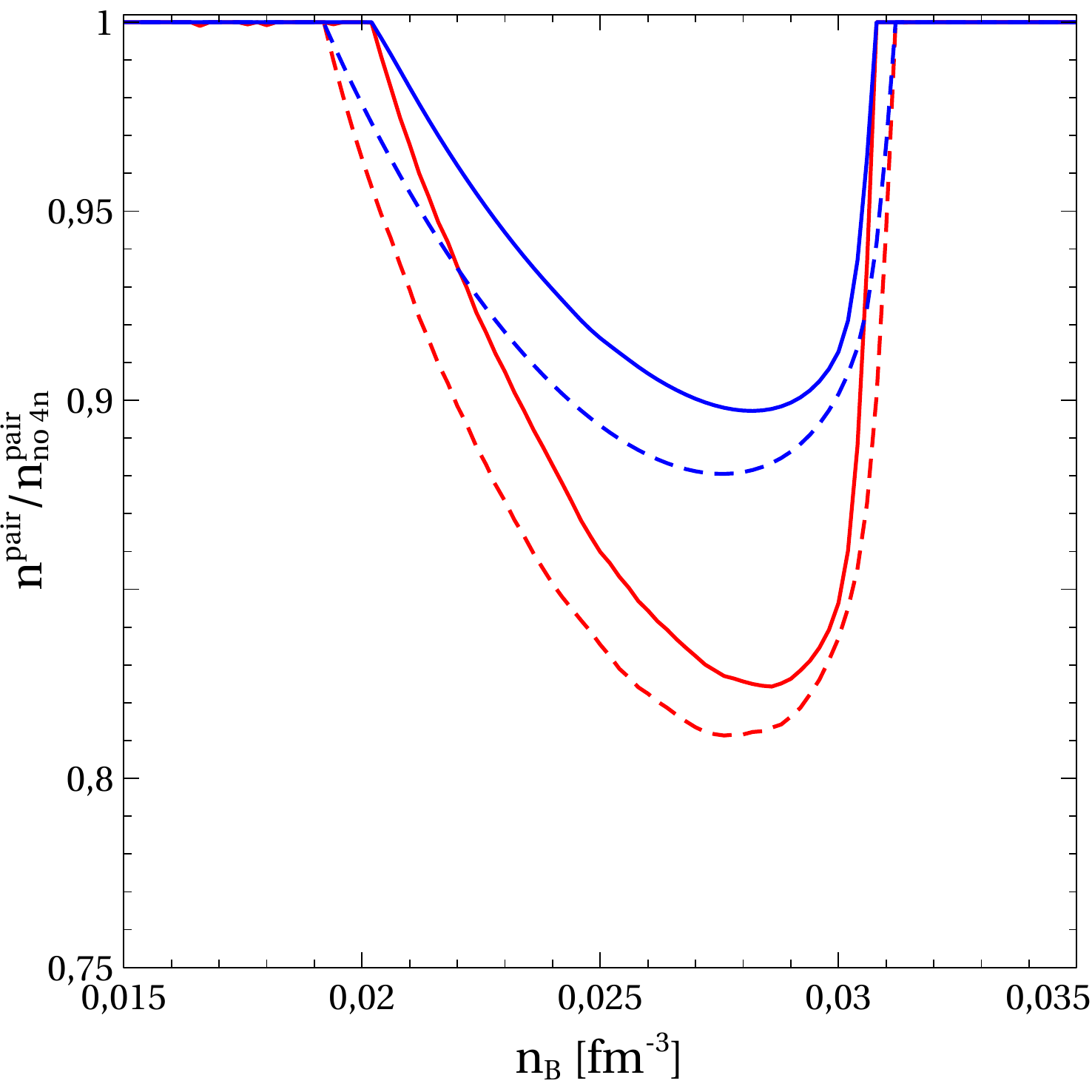}
\includegraphics[width=0.85\columnwidth]{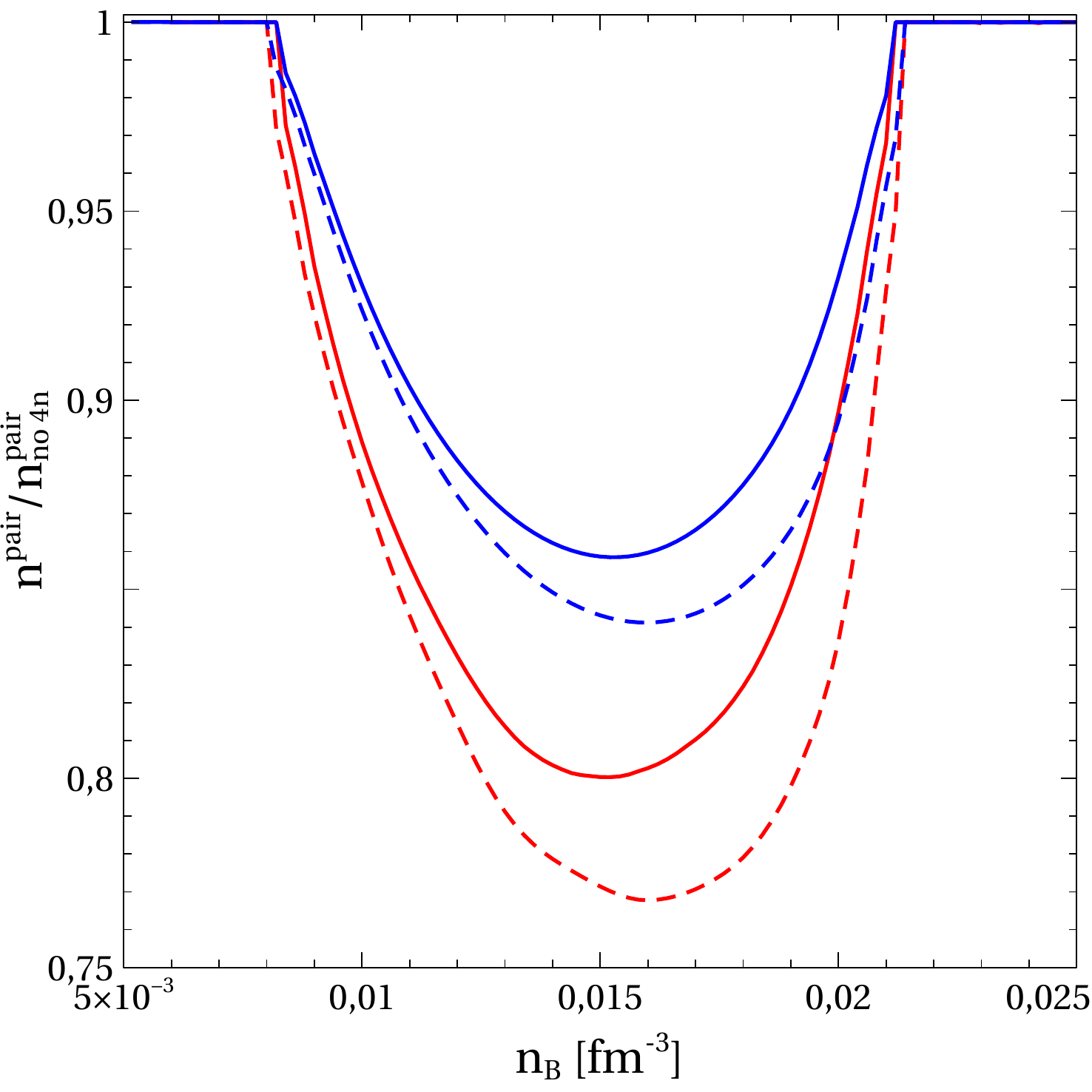}
\caption{Ratio of densities of paired nucleons (neutrons in blue, protons in red) in presence and in absence of tetraneutrons $n^{\rm pair}/n^{\rm pair}_{\rm no~4n}$ as a function of baryon density calculated for sets  A (upper panel) and  B (lower panel). Finite volume corrections (solid line) and pointlike approximations  (dashed line) are also shown.}
\label{fig3}
\end{figure}
The behaviour of the neutron pairing gap  as a function of momentum is shown in Fig. \ref{fig2} for  model set A (upper panel) and set  B (lower  panel). The baryonic density is  fixed at $n_B= 0.0275$ $\rm fm^{-3}$ for set A  and $n_B= 0.0150$ $\rm fm^{-3}$ for set  B, close to the maximum of the baryonic charge fraction for tetraneutrons depicted in Fig. \ref{fig12}. We find that the corresponding gap for protons is much  smaller, in agreement with standard calculations \cite{pgap} and is not shown. We depict the cases of pointlike (thin dashed line) and finite-size tetraneutrons with $R=5$ fm (thin solid line). The case with no tetraneutrons is shown with thick solid line and for that case neutron Fermi momentum is  $k_F = 0.9323$ $\rm fm^{-1}$ (set A) and $k_F = 0.7625$ $\rm fm^{-1}$ (set B). When tetraneutrons are present set B yields more pronounced differences at low momentum. This, in turn, translates into the pairing fractions as we will later see in the manuscript. We consider $\Gamma_{4n}$ values as they appear in Table \ref{table2}. We can see that the main difference arises for the low momentum when the BEC manifests more clearly the difference among paired and unpaired neutrons. When tetraneutrons are present the amplitude is somewhat decreased with respect to the case without them. In addition, the dependence on the coupling ratio $x_{4n\sigma}$ along with the decay width is nevertheless weak leading to slight changes in the gap profile. Note that for values of $\Gamma_{4n}$ outside the allowed bands (see Fig. \ref{fig11}) no solution with tetraneutrons would exist.

Pairing of nucleons is controlled by the gap $\Delta_N$, which is strongly influenced by the tetraneutron BEC.  Therefore, this condensate also significantly changes the density of paired particles. For nucleons it reads

\footnotesize
\begin{eqnarray}
\label{XVI}
n^{\rm pair}_N=\int\limits_0^\infty \frac{dk'}{2\pi^2}
\left(1-\frac{|\epsilon_N(k',m^*_N)-\mu^*_N|}
{\sqrt{(\epsilon_N(k',m^*_N)-\mu^*_N)^2+\Delta^2_N(k')}}\right).
\end{eqnarray}
\normalsize
We show in Fig. \ref{fig3} the behaviour of the ratios of densities of paired nucleons in presence and in absence of tetraneutrons $n^{\rm pair}/n^{\rm pair}_{\rm no~4n}$ as a function of the baryonic density for model sets A (upper panel) and B (lower panel). The cases of finite-size (solid line) and pointlike tetraneutrons (dashed line) are also shown. Clearly, before the tetraneutron onset densities  and beyond their dissolution ones this fraction has a unit value. The situation changes once the onset density for formation of the tetraneutron condensate is reached (see actual values in Table \ref{table2}) as there is a pronounced decrease for both nucleon types. As it is  seen, this reduction is larger for protons at densities belonging to the region of the outer NS crust. Although our model only considers light clusters it is worth noting that the addtional pressence of a fraction of heavier species, nuclei, is to be considered in future works as it could lead to a supression of the light bound clusters as found in \cite{wu}.

\section{Conclusions}
We have explored the possible condensation of tetraneutrons in neutron rich matter inside Neutron Stars. We assumed that they can be produced in a  thermodynamically equilibrated medium whose properties are controlled by the corresponding chemical potentials. As a first step, neglecting  higher order correlations,  we started by using a relativistic density functional approach and we find that scanning a prescribed range of couplings of tetraneutrons  to the $\sigma$, $ \omega,\rho$ fields based on arguments of isospin symmetry, similar to those used in the literature for other clusters, their decay width largely determines the actual  presence of a  tetraneutron condensate. If that was the case it can lead to a partial suppression of the S-wave nucleon pairing manifested through a reduction of the fraction of paired protons and neutrons in the system. This happens due to a more energetically favourable combination of neutrons to tetraneutrons condensing to the lowest energy state. Pauli blocking effects have been partially included using an effective treatment in the same fashion already used for stable nuclear clusters. In our model, the fraction of tetraneutrons depends on their actual decay width and, if allowed, it remains small (up to $4\%$ of baryonic density)  restricted to densities about one tenth of nuclear saturation density, thus we expect that  they will have a very mild impact on the equation of state of dense nuclear matter or NS masses. We expect, however, that it could most likely affect the microscopic behaviour of the neutron rich matter in the crust. Further work is needed to clarify this latter aspect and it is left for future contributions.\\

\section*{Acknowledgements} 
We thank valulable discussions with  F. Gulminelli, C. Providencia and A. Valcarce. 
This work was performed with financial support from the project SA083P17 by Junta de Castilla y Le\'on and University of Salamanca. We also acknowledge CA16214 (PHAROS) and the Spanish Red Consolider MultiDark FPA2017-90566-REDC. The authors would like to thank FCSCL (Fundaci\'on Centro de Supercomputaci\'on de Castilla y Le\'on) for providing access to a cluster of its supercomputer Cal\'endula.



\end{document}